\documentstyle[12pt,epsf,axodraw]{article} \textwidth 16.5cm
\newcommand{\beq}{\begin{equation}}
\newcommand{\eeq}{\end{equation}}
\newcommand{\bea}{\begin{eqnarray}}
\newcommand{\eea}{\end{eqnarray}}

\begin{document}
\begin{titlepage}

\begin{center}
{\LARGE \bf BFKL resummation effects in the $\gamma^* \gamma^*$ total hadronic
cross section}
\end{center}

\vskip 0.5cm

\centerline{F. Caporale$^{1\dag}$, D.Yu. Ivanov$^{2\P}$ and
A.~Papa$^{1\ddagger}$}

\vskip .6cm

\centerline{${}^1$ {\sl Dipartimento di Fisica, Universit\`a della Calabria,}}
\centerline{\sl Istituto Nazionale di Fisica Nucleare, Gruppo collegato di
Cosenza,}
\centerline{\sl Arcavacata di Rende, I-87036 Cosenza, Italy}

\vskip .2cm

\centerline{${}^2$ {\sl Sobolev Institute of Mathematics,}}
\centerline{\sl 630090 Novosibirsk, Russia}

\vskip 2cm

\begin{abstract}
We study in the BFKL approach the total hadronic cross section for the
collision of two virtual photons for energies in the range of LEP2 and in the
range of future linear colliders.
The BFKL resummation is done at the next-to-leading order in the BFKL Green's
function; photon impact factors are taken instead at the leading order, but
with the inclusion of the subleading terms required by invariance under
changes of the renormalization scale and of the BFKL scale $s_0$. We compare
our results with previous estimates based on a similar kind of approximation.
\end{abstract}


$
\begin{array}{ll} ^{\dag}\mbox{{\it e-mail address:}} &
\mbox{caporale@cs.infn.it}\\
^{\P}\mbox{{\it e-mail address:}} &
\mbox{d-ivanov@math.nsc.ru}\\
^{\ddagger}\mbox{{\it e-mail address:}} &
\mbox{papa@cs.infn.it}\\
\end{array}
$

\end{titlepage}

\vfill \eject

\section{Introduction}

The total hadronic cross section for the collision of two off-shell photons
with large virtualities is a fundamental observable, since, similarly to the
process of $e^+e^-$ annihilation into hadrons, is fully under control of
perturbative QCD. In fixed order calculations, the dominant
contribution at low energies comes from the quark box, calculated at the
leading-order (LO) in Ref.~\cite{Budnev:1974de} (see also
Ref.~\cite{Schienbein:2002wj}) and at the next-to-LO (NLO) in
Ref.~\cite{Cacciari:2000cb}. In Ref.~\cite{BL03} the resummation of double
logs appearing in the NLO corrections to the quark box has been also studied.
At higher energies the diagrams with gluon exchange in the
$t$-channel become more important since they have different power asymptotics
for $s\to \infty$ in comparison to the $t$-channel quark exchange;
at higher orders such contributions contain powers of single logarithms
of energy which can be resumed in the frame of the BFKL approach.

It is widely believed that the $\gamma^*\gamma^*$ total cross section is the
best place for the possible manifestation of the BFKL dynamics~\cite{BFKL}
at the energies of future linear colliders (for a review, see
Ref.~\cite{Wallon:2007xc}). For this reason, many
papers~\cite{photons_BFKL} have considered the inclusion of the BFKL
resummation of leading energy logarithms.  In a remarkable
paper~\cite{Brodsky:2002ka} (see also Ref.~\cite{Brodsky:1998kn}),
BFKL resummation effects have been taken into account also at the subleading
order and evidence has been presented that the appearance of BFKL dynamics is
compatible with experimental data already at the energies of
LEP2~\cite{Achard:2001kr,Abbiendi:2001tv}.

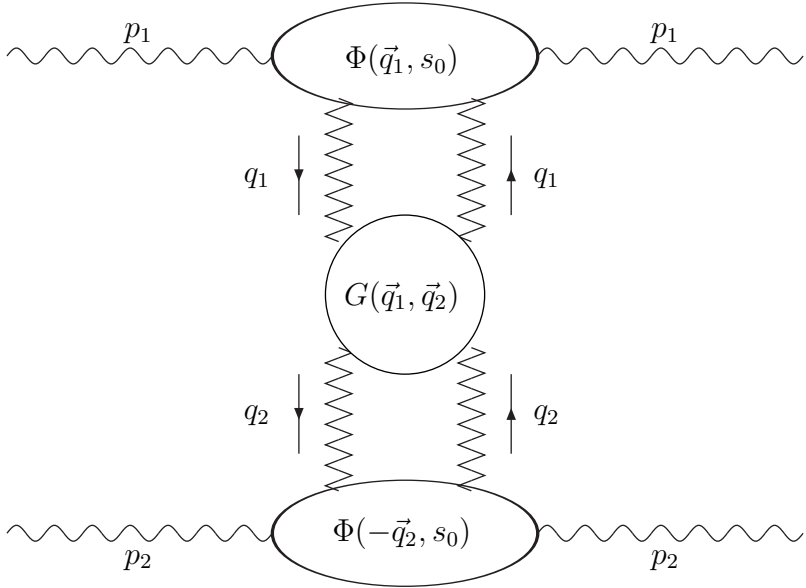
\begin{figure}[tb]
\centering
\setlength{\unitlength}{0.35mm}
\begin{picture}(300,200)(0,0)

\Photon(0,190)(100,190){3}{7}
\Photon(200,190)(300,190){3}{7}
\Text(50,200)[c]{$p_1$}
\Text(250,200)[c]{$p_1$}
\Text(150,190)[]{$\Phi(\vec q_1, s_0)$}
\Oval(150,190)(20,50)(0)

\ZigZag(125,174)(125,120){5}{7}
\ZigZag(175,174)(175,120){5}{7}
\ZigZag(125,26)(125,80){5}{7}
\ZigZag(175,26)(175,80){5}{7}

\ArrowLine(110,160)(110,130)
\ArrowLine(190,130)(190,160)
\ArrowLine(110,70)(110,40)
\ArrowLine(190,40)(190,70)

\Text(100,145)[r]{$q_1$}
\Text(200,145)[l]{$q_1$}
\Text(100,55)[r]{$q_2$}
\Text(200,55)[l]{$q_2$}

\GCirc(150,100){30}{1}
\Text(150,100)[]{$G(\vec q_1,\vec q_2)$}

\Photon(0,10)(100,10){3}{7}
\Photon(200,10)(300,10){3}{7}
\Text(50,0)[c]{$p_2$}
\Text(250,0)[c]{$p_2$}
\Text(150,10)[]{$\Phi(-\vec q_2,s_0)$}
\Oval(150,10)(20,50)(0)

\end{picture}

\caption[]{Schematic representation of the elastic amplitude for the
$\gamma^*(p_1)\, \gamma^*(p_2)$ forward scattering.}
\label{fig:BFKL}
\end{figure}

In the BFKL approach, both in the leading logarithmic approximation (LLA),
which means resummation of leading energy logarithms, all terms
$(\alpha_s\ln(s))^n$, and in the next-to-leading approximation (NLA), which
means resummation of all terms $\alpha_s(\alpha_s\ln(s))^n$, the imaginary part
of the amplitude for a large-$s$ hard collision process can be written as
the convolution of the Green's function of two interacting Reggeized gluons
with the impact factors of the colliding particles (see, for example,
Fig.~\ref{fig:BFKL}).

The Green's function is determined through the BFKL equation and is
process-inde\-pen\-dent. The NLO kernel of the BFKL equation for singlet color
representation in the $t$-channel and for forward scattering, relevant for
the determination of a total cross section, has been achieved in
Refs.~\cite{NLA-kernel}, after the long program of calculation
of the NLO corrections~\cite{NLA-corrections} (for a review, see
Ref.~\cite{news}).

The other essential ingredient to build up the $\gamma^* \gamma^*$ total cross
section is the impact factor for the virtual photon to virtual photon
transition.
Its calculation in the NLO is rather complicated and has been completed only
after year-long efforts~\cite{gammaIF}. The result, obtained in the momentum
representation, is known to a large extent in numerical form. After gathering
the NLO virtual photon impact factors with the NLO BFKL Green's function, the
prediction for the energy dependence of the NLA $\gamma^* \gamma^*$ total cross
section will become available. This remaining step is, however, rather
difficult, so it may be interesting in the meanwhile to get an estimate of NLA 
BFKL effects, using an approximated procedure and possibly refining previous
analysis of the same kind.

In this paper we estimate the energy dependence of the $\gamma^* \gamma^*$
total cross total hadronic in an energy range which covers
LEP2 and future linear colliders. The procedure we follow is approximate, since
we use the singlet forward NLO BFKL Green's function together with forward
$\gamma^* \to \gamma^*$ impact factors {\em at the leading order}.
However, in the impact factors we include the subleading terms required by the
invariance of the full amplitude at the NLA under change of the renormalization
scale and of the energy scale $s_0$ entering the BFKL approach. The neglect
of other subleading corrections to the impact factor certainly affects the low
energy behavior of the cross section, but should not spoil the high energy
regime. A more detailed discussion on this point will be presented later on.

The calculation goes along the same steps as for the amplitude $\gamma^*
\gamma^* \to V V$, with $V=\rho^0, \omega, \phi$ a light neutral vector meson.
In that case, however, the relevant impact factor, namely the $\gamma^* \to V$
impact factor, was available in closed analytic form in the NLO, up to
contributions suppressed as inverse powers of the photon
virtuality~\cite{IKP04}. Therefore, the amplitude could be evaluated fully in
the NLA~\cite{mesons_1-2,mesons_3}, previous estimations being based on fixed
perturbative order calculations~\cite{Pire:2005ic} and on partial inclusion of
NLA BFKL effects~\cite{Enberg:2005eq} \footnote{See
also Ref.~\cite{Pire:2006ik} for an analysis of the QCD factorization
properties of this amplitude.}. As in Refs.~\cite{mesons_1-2,mesons_3}, the
convolution between the impact factors of the colliding photons, taken with
equal virtualities, and the BFKL Green's function is performed in the space
conjugated to the transverse momentum space, namely in the so-called
$\nu$-space or, equivalently, through the spectral decomposition on the
eigenfunctions of the LO BFKL kernel. Similarly to
Refs.~\cite{mesons_1-2,mesons_3} the large NLA corrections are handled
by the adoption of suitable optimization methods of the perturbative series; in
particular, we used the principle of minimal sensitivity (PMS)~\cite{Stevenson}
and the the Brodsky-Lepage-Mackenzie (BLM) method~\cite{BLM}.

The approximation of using LO impact factors in combination with NLO BFKL
Green's function is not new. It has been exploited already in
in Ref.~\cite{Brodsky:2002ka} for the $\gamma^* \gamma^*$ total cross
section, in Ref.~\cite{Enberg:2005eq} for the $\gamma^* \gamma^*
\to VV$ amplitude, in Ref.~\cite{Vera:2007kn} for the production of 
Mueller-Navelet jets at hadron colliders and in Refs.~\cite{Vera:2007dr,
Kepka:2006xe} for the production of forward jets in deep-inelastic-scattering.
In comparison with Ref.~\cite{Brodsky:2002ka}, in the
present paper, the elements of novelty are the following:
\begin{itemize}
\item the optimization procedures to stabilize the perturbative series are
performed on the amplitude itself and not on the NLO Pomeron intercept; we
believe that this is more natural since a perturbative intercept is not a
physical quantity;
\item the impact factors, although taken at the LO, contain the appropriate
NLO terms, so that the dependence on the energy scales entering the process
(the renormalization scale $\mu_R$ and the parameter $s_0$ introduced in the
BFKL approach) is pushed to the next-to-NLA; this makes the effect of $s_0$
on the numerical result less pronounced than in Ref.~\cite{Brodsky:2002ka};
moreover, in our approach the value of $s_0$ (as well as that of $\mu_R$) is
determined by the optimization procedure and is not a free parameter;
\item two optimization methods are used, thus having a control of systematic
effects at work.
\end{itemize}

The paper is organized as follows: in the next Section we give the expression
of the cross section; in Section~3
we present the numerical results; in Section~4 we draw our conclusions.

\section{The $\gamma^* \gamma^*$ total cross section}

The total hadronic cross section of two unpolarized photons with virtualities
$Q_1$ and $Q_2$ can be obtained from the imaginary part of the forward
amplitude (see Fig~\ref{fig:BFKL}) and within LO BFKL is given by the following
expression (see, for instance, Ref.~\cite{Brodsky:2002ka}):
\begin{equation}
\sigma^{\gamma^{*} \gamma^{*}}_{tot} (s,Q_1,Q_2) =
\sum_{i,k=T,L} \frac{1}{(2 \pi)^2 Q_1 Q_2}
\int\limits^{+\infty}_{-\infty} d\nu \left(\frac{Q_1^2}{Q_2^2}\right)^{i\nu}
F_i(\nu) F_k(-\nu) \left(\frac{s}{s_0}\right)^{\bar \alpha_s \chi(\nu)} \; ,
\label{sigmaLO}
\end{equation}
where $\bar \alpha_s\equiv \alpha_s N_c/\pi$, $\chi(\nu)$ is the so-called
characteristic BFKL function,
\beq
\chi(\nu)=2\psi(1)-\psi\left(\frac{1}{2}+i\nu\right)-\psi\left(\frac{1}{2}
-i\nu\right)\;,
\eeq
and
\beq
F_T(\nu) = F_T(-\nu)= \alpha \, \alpha_s \left( \sum_q e_q^2 \right)
\frac{\pi}{2} \frac{\left(\frac{3}{2} - i\nu\right)\left(\frac{3}{2}
+ i\nu\right)
\Gamma\left(\frac{1}{2} - i\nu\right)^2
\Gamma\left(\frac{1}{2} + i\nu\right)^2}{\Gamma(2-i\nu) \Gamma(2+i\nu)} \, ,
\eeq
\beq
F_L(\nu) = F_L(-\nu)= \alpha \, \alpha_s \left( \sum_q e_q^2 \right) \pi
\frac{\Gamma\left(\frac{3}{2} - i\nu\right)\Gamma\left(\frac{3}{2} + i\nu
\right)\Gamma(\frac{1}{2} - i\nu) \Gamma(\frac{1}{2} + i\nu)}
{\Gamma(2-i\nu) \Gamma(2+i\nu)}
\eeq
are the LO impact factors for transverse and longitudinal polarizations,
respectively. In the previous equations, $\alpha$ is the electromagnetic
coupling constant, the summation extends over all active quarks (taken
massless) and $e_q$ is the quark electric charge in units of the electron
charge. In the expression (\ref{sigmaLO}) for the LO BFKL cross section the
argument of the strong and electromagnetic coupling constants and the value of
the scale $s_0$ are not fixed.

Following the procedure of Refs.~\cite{mesons_1-2}, it is possible
to write down the cross section with the inclusion of NLO corrections in the
Green's function only, while keeping the impact factors at the LO:
\beq
\sigma^{\gamma^{*} \gamma^{*}}_{tot} (s,Q_1,Q_2)
= \frac{1}{(2 \pi)^2 Q_1 Q_2}
\int\limits^{+\infty}_{-\infty} d\nu \left(\frac{Q_1^2}{Q_2^2}\right)^{i\nu}
\left(\frac{s}{s_0}\right)^{\bar \alpha_s(\mu_R) \chi(\nu)}
\sum_{i,k=T,L}  F_i(\nu) F_k(-\nu)
\eeq
\[
\times
\left\{ 1 + \bar \alpha_s^2(\mu_R) \ln\left(\frac{s}{s_0}\right) \left[ \bar
\chi(\nu)+\frac{\beta_0}{8N_c}\chi(\nu)\left(-\chi(\nu)+\frac{10}{3}
+ 2\ln\frac{\mu_R^2}{Q_1Q_2} \right) \right] \right\} \; ,
\]
where
\bea
\bar\chi(\nu)\,&=&\,-\frac{1}{4}\left[\frac{\pi^2-4}{3}\chi(\nu)-6\zeta(3)-
\chi^{\prime\prime}(\nu)-\frac{\pi^3}{\cosh(\pi\nu)}
\right.
\nonumber \\
&+& \left.
\frac{\pi^2\sinh(\pi\nu)}{2\,\nu\, \cosh^2(\pi\nu)}
\left(
3+\left(1+\frac{n_f}{N_c^3}\right)\frac{11+12\nu^2}{16(1+\nu^2)}
\right)
+\,4\,\phi(\nu)
\right] \, ,
\eea
\beq
\phi(\nu)\,=\,2\int\limits_0^1dx\,\frac{\cos(\nu\ln(x))}{(1+x)\sqrt{x}}
\left[\frac{\pi^2}{6}-\mbox{Li}_2(x)\right]\, , \;\;\;\;\;
\mbox{Li}_2(x)=-\int\limits_0^xdt\,\frac{\ln(1-t)}{t} \, .
\eeq

In fact, the requirement of invariance of the amplitude at the NLA under
renormalization group transformation and under change of the energy scale
$s_0$ allows to fix the $\mu_R$- and $s_0$-dependent terms in the NLO impact
factors, so that we can get the following expression for the total cross
section, given by
\begin{eqnarray}
\sigma^{\gamma^{*} \gamma^{*}}_{tot} (s,Q_1,Q_2)
&=& \frac{1}{(2 \pi)^2 Q_1 Q_2}
\int\limits^{+\infty}_{-\infty} d\nu \left(\frac{Q_1^2}{Q_2^2}\right)^{i\nu}
\left(\frac{s}{s_0}\right)^{\bar \alpha_s(\mu_R) \chi(\nu)}
\sum_{i,k=T,L} F_i(\nu) F_k(-\nu)  \nonumber \\
&\times&
\left\{ 1 +  \bar \alpha_s(\mu_R) A(s_0) +  \bar \alpha_s(\mu_R)B(\mu_R)
+ \bar \alpha_s^2(\mu_R) \ln\left(\frac{s}{s_0}\right)
\Bigg[ \bar \chi(\nu)
\right. \nonumber \\
&+& \left. \left.
\frac{\beta_0}{8N_c}\chi(\nu)\left(-\chi(\nu)+\frac{10}{3}
+ 2\ln\frac{\mu_R^2}{Q_1 Q_2} \right) \right] \right\} \;,
\label{sigma}
\end{eqnarray}
with
\begin{equation}
A(s_0) = \chi(\nu) \ln\frac{s_0}{Q_1 Q_2}\;,
\;\;\;\;\;\;\;\;\;
B(\mu_R)= \frac{\beta_0}{2N_c} \ln\frac{\mu_R^2}{Q_1 Q_2}\;.
\label{AB}
\end{equation}

The above expression for the cross section can be conveniently represented
as a series,
\begin{equation}
Q_1 Q_2 \, \sigma^{\gamma^{*} \gamma^{*}}_{tot}  =
\frac{1}{(2 \pi)^2} \left\{ b_0 + \sum_{n=1}^{\infty}\bar \alpha_s(\mu_R)^n
\, b_n \, \biggl[\ln^n \frac{s}{s_0} + d_n(s_0,\mu_R)\ln^{n-1}\frac{s}{s_0}
\biggr]\right\} \;,
\end{equation}
with coefficients
\begin{equation}
b_n=\int\limits^{+\infty}_{-\infty}d\nu \left( \frac{Q_1^2}{Q_2^2} \right)^{i
\nu} \: \sum_{i,k=T,L} F_i(\nu) F_k(-\nu) \: \frac{\chi^n(\nu)}{n!} \;,
\end{equation}
determined by the kernel and the impact factors in LLA, and
\[
d_n=n\ln\frac{s_0}{Q_1Q_2}+\frac{\beta_0}{4N_c}
\left[\frac{b_{n-1}}{b_n} \left((n+1)\ln\frac{\mu_R^2}{Q_1Q_2}+ \frac{5}{3}
(n-1) \right)-\frac{n(n-1)}{2} \right]
\]
\begin{equation}
+ \frac{1}{b_n} \int\limits^{+\infty}_{-\infty}d\nu
\left(\frac{Q_1^2}{Q_2^2}\right)^{i \nu} \sum_{i,k=T,L} F_i(\nu) F_k(-\nu)
\; \frac{\chi^{n-2}(\nu)}{(n-2)!}\: \bar \chi(\nu)\;,
\end{equation}
determined by the NLO corrections. The series representation is one of the
infinitely many possible ways, equivalent with NLA accuracy, to represent the
total cross section. It has the advantage to make manifest the BFKL resummation
of leading and subleading energy logarithms and is very practical in numerical
computations. The well-known feature of the large NLA BFKL corrections is
revealed by being the $d_n$ coefficients of opposite sign with respect to the
$b_n$ and increasing with $n$ in absolute value.

The neglect of NLO corrections, except for $\mu_R$- and $s_0$-dependent terms,
to the impact factors affects the value of the $d_n$ coefficients. In the case
of the $\gamma^*\gamma^*\to VV$ process, it turned out that the contribution
to $d_n$ from the kernel starts to dominate over that from the impact factors
for $n\geq 4$. This makes evident the fact that the high-energy behavior of
the amplitude is weakly affected by the NLO corrections to the impact factor.
Therefore, our approximated $\gamma^*\gamma^*$ total cross section should
compare better and better with the correct result as the energy increases.

\begin{figure}[ht]
\begin{minipage}{70mm}
\begin{center}
\setlength{\unitlength}{0.35mm}
\begin{picture}(150,150)(0,0)
\Photon(10,140)(50,100){3}{7}
\ArrowLine(50,100)(100,100)
\Photon(100,100)(140,140){3}{7}
\ArrowLine(50,100)(50,50)
\Photon(10,10)(50,50){3}{7}
\ArrowLine(50,50)(100,50)
\Photon(100,50)(140,10){3}{7}
\ArrowLine(100,50)(100,100)
\end{picture}
\end{center}
\end{minipage}
\begin{minipage}{90mm}
\begin{center}
\setlength{\unitlength}{0.35mm}
\begin{picture}(150,150)(0,0)
\Photon(10,140)(50,100){3}{7}
\Line(50,100)(100,50)
\Photon(100,100)(140,140){3}{7}
\ArrowLine(50,100)(50,50)
\Photon(10,10)(50,50){3}{7}
\Line(50,50)(100,100)
\Photon(100,50)(140,10){3}{7}
\ArrowLine(100,50)(100,100)
\end{picture}
\end{center}
\end{minipage}
\caption[]{Quark box LO diagrams.}
\label{qbox}
\end{figure}
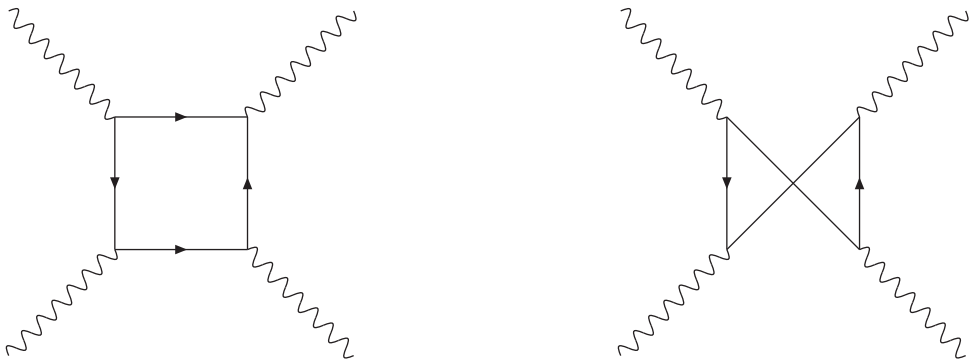

Our aim in this paper is to study the dependence of the cross section given in
Eqs.~(\ref{sigma})-(\ref{AB}) on the center-of-mass energy, both in the range
of LEP2, where experimental data are
available~\cite{Achard:2001kr,Abbiendi:2001tv}, and
of the future linear colliders. In order to stabilize the perturbative
series, it is necessary to resort to some optimization procedure, exploiting
the freedom to vary the energy parameters, $\mu_R$ and $s_0$, without
corrupting the calculation but at the next-to-NLA. Following
Refs.~\cite{mesons_1-2}, we use both
the principle of minimal sensitivity (PMS method)~\cite{Stevenson} and the
Brodsky-Lepage-Mackenzie (BLM) method~\cite{BLM}: for some selected values of
the energy $s$ in the region of interest the optimal scales $\mu_R$ and $s_0$
are found and the cross section is thus determined. Then, the curve giving the
cross section {\it vs.} the energy is obtained by interpolation.

In order to compare the theoretical prediction with the existing data from
LEP2, we cannot neglect the contribution from LO quark box diagrams shown in
Fig.~\ref{qbox}, which is of order $\alpha^2 (\ln s)/s$. We take this
contribution from Ref.~\cite{Budnev:1974de}. To be definite, we take the
so-called $\phi$-averaged $\gamma^*\gamma^*$ total (hadronic) cross section,
written as
\beq
\sigma^{\gamma^{*} \gamma^{*}}_{QBOX} (s,Q_1,Q_2) = \sum_q e_q^4
\biggl(\sigma_{TT} +2\sigma_{TS}+\sigma_{SS}\biggr)\;,
\label{sigmaQBOX}
\eeq
with the $\sigma_{ik}$, $i,k=T,S$ given in the Appendix~E of
Ref.~\cite{Budnev:1974de} and evaluated for massless $u$-, $d$-, $c$-,
$s$-quarks and massive $b$- and $t$-quarks.
On the other hand, the soft Pomeron contribution,
if estimated within the vector-dominance model, is proportional to
$\sigma_{\gamma^*\gamma^*} \sim (m^2_V/Q^2)^4 \sigma_{\gamma \gamma}$ and
is therefore suppressed for highly virtual photons. In the following analysis
we neglect such higher twist contributions.

\section{Numerical results}

We restrict ourselves to the case of symmetric kinematics, which means equal
virtuality $Q_1 =Q_2 \equiv Q$ for the two photons. This is the so-called
``pure BFKL regime'', as opposite to the ``DGLAP regime'' realized for strongly
ordered photon virtualities.

Let us start with the PMS optimization method~\cite{Stevenson}. Setting
$Y\equiv\ln(s/Q^2)$ and $Y_0\equiv\ln(s_0/Q^2)$, we require that, for each
value of $Y$, the cross section given in Eqs.~(\ref{sigma})-(\ref{AB}) is the
least sensitive to the variation of the scales $\mu_R$ and $Y_0$.
For a first analysis we choose $Q^2$=17~GeV$^2$ ($n_f$=4), in order to
compare with the experimental data from CERN LEP2 collected for
$<Q^2>$=16~GeV$^2$ (L3) and $<Q^2>$=18~GeV$^2$ (OPAL). We have found that the
cross section is quite stable under variation of the two scales and generally
exhibits only one stationary point (local maximum). In the few cases when the
cross section presented more than one stationary point, we choose as optimal
parameters those corresponding to the smoothest stationary point. The typical
values that optimize the cross section turned out to be $\mu_R \simeq 3Q$ and
$Y_0 \simeq 2$. Note that for the $\gamma^*\gamma^* \to VV$ amplitude at
$Q^2$=24~GeV$^2$ ($n_f$=5) the same procedure led to optimal values for
$\mu_R$ as large as $\sim 10Q$.

The other optimization procedure we considered is inspired by the BLM
method~\cite{BLM}: we perform a finite renormalization to the momentum (MOM)
scheme with $\xi=0$ and then choose the renormalization scale in order to
remove the $\beta_0$-dependent part in the cross section. The renormalization
is defined as follows:
\[
\alpha_s \to \alpha_s \left[ 1+ T_{MOM} (\xi = 0)\frac{\alpha_s}{\pi} \right]
\;,
\;\;\;\; T_{MOM} (\xi = 0) = T^{conf}_{MOM} + T^{\beta}_{MOM} \;,
\]
\[
T^{conf}_{MOM}= \frac{N_c}{8}\frac{17}{2} I \;, \;\;\;\; T^{\beta}_{MOM}
= -\frac{\beta_0}{2} \left[1 + \frac{2}{3}I \right] \;, \;\;\;\; I
\simeq 2.3439 \; .
\]
Then, following~\cite{mesons_1-2}, for each value of $Y$ we found the pairs
$(Y_0,\mu_R)$ for which the term proportional to $\beta_0$ in the renormalized
cross section vanishes. Then, among the resulting pairs, we determined the
optimal one according to the PMS principle. The typical values of the scales
found in this way are very similar to those obtained with the other method.
We stress that this way of using the BLM optimization method is somewhat
different from Ref.~\cite{Brodsky:2002ka}, since there the $\gamma^*\gamma^*$
total cross section was built using a Pomeron intercept optimized by the BLM
method (see Ref.~\cite{Brodsky:1998kn}). Here we apply the BLM optimization
procedure to the cross section itself, which is a well-defined physical
quantity, while the perturbative Pomeron intercept can not be derived
directly from experiment data.

\begin{figure}[tb]
\centering
\hspace{-1cm}
{\epsfysize 8cm \epsffile{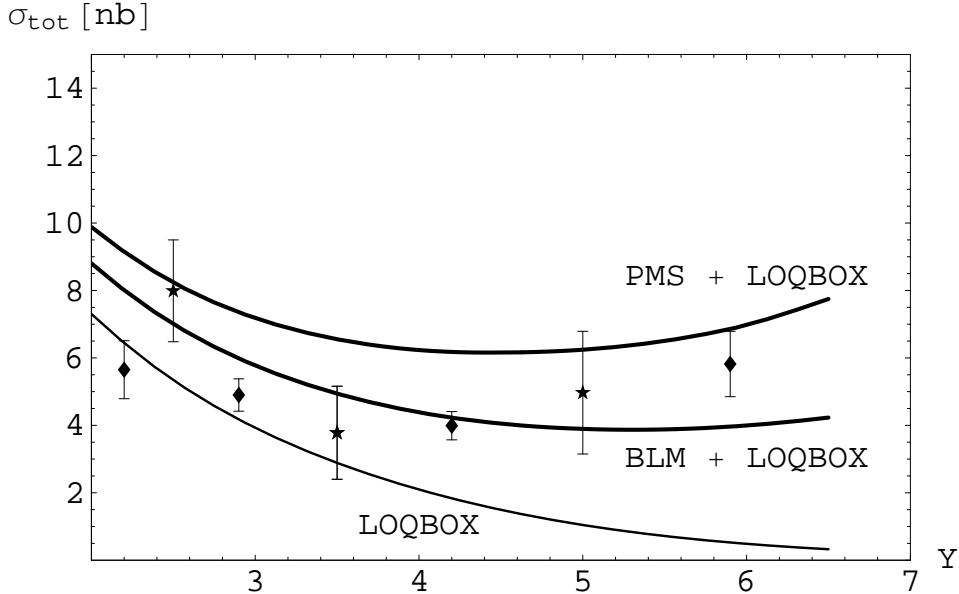}}
\caption[]{Energy dependence of the total cross section for the collision of
two photons with virtualities $Q^2$=17~GeV$^2$ ($n_f$=4) as predicted by the
PMS and BLM methods, with the inclusion of the LO quark box contribution. For
comparison, experimental data from OPAL~\cite{Abbiendi:2001tv}
(stars, $Q^2$=18~GeV$^2$) and L3~\cite{Achard:2001kr} (diamonds,
$Q^2$=16~GeV$^2$) are shown.}
\label{fig:sigma}
\end{figure}

In this work we consider two regions of energy: the CERN LEP2 region and a
higher energy region, possibly reachable in future linear colliders.
In the first of these regions we can compare
our results with LEP2 experimental data and with the determinations of
Ref.~\cite{Brodsky:2002ka}. We admit that in this region the neglect of NLO
corrections to the impact factors can play an important role; since these
corrections are negative (see, for instance, Ref.~\cite{Chachamis:2006zz}),
our prediction will certainly overestimate the true NLO result. In the second
energy region considered, we expect the role of NLO corrections to the impact
factor be less relevant and, therefore, our prediction closer to the true NLA
BFKL result.

In Fig.~\ref{fig:sigma} we summarize our results for the CERN LEP2 region:
we show the NLA BFKL curves obtained
by the PMS and the BLM methods, to which we added the contribution of the LO
quark box diagrams. For comparison we put in this plot also the experimental
data from CERN LEP2, namely three data points from OPAL~\cite{Abbiendi:2001tv}
($Q^2$=18~GeV$^2$)
and four data points from L3~\cite{Achard:2001kr} ($Q^2$=16~GeV$^2$).
We observe first of all that the difference between the two theoretical
curves can be taken as an estimate of the systematics effects which underlay
the optimization procedures adopted here. The fact that the PMS curve is
systematically above the BLM curve is not surprising, since the stationary
point for the amplitude in the space of the parameters $Y_0$ and $\mu_R$ is
always a local maximum, sometimes an absolute maximum, for varying $Y$.
The comparison with experiments shows that the PMS curve overestimates data,
while the BLM curve seems to be more in agreement with them. However, if we
recall that in both cases a negative contribution from NLO impact factors is
being missed, it seems that the PMS curve has better chances to agree with data
in the fully NLA BFKL calculation. Around the energy for which the
applicability condition for the BFKL resummation, $\bar \alpha_s Y \sim 1$,
is satisfied, which in the considered kinematics
corresponds to $Y\sim 5$, {\em both} the PMS and BLM curves agree with
experimental data within the (large) errors. Finally, we remark that the
determination from
Ref.~\cite{Brodsky:2002ka} falls between our two curves from PMS and BLM
methods.
It is also important to note the the high-energy rise of the data for the cross
section cannot be described only by the LO quark-box, nor can it be explained
by the NLO quark box contribution~\cite{Cacciari:2000cb} (the L3 datum at
$Y=6$ is underestimated by $\sim$4 standard deviations, see
Ref.~\cite{Brodsky:2002ka}).

\begin{figure}[ht]
\centering
\hspace{-1cm}
{\epsfysize 8cm \epsffile{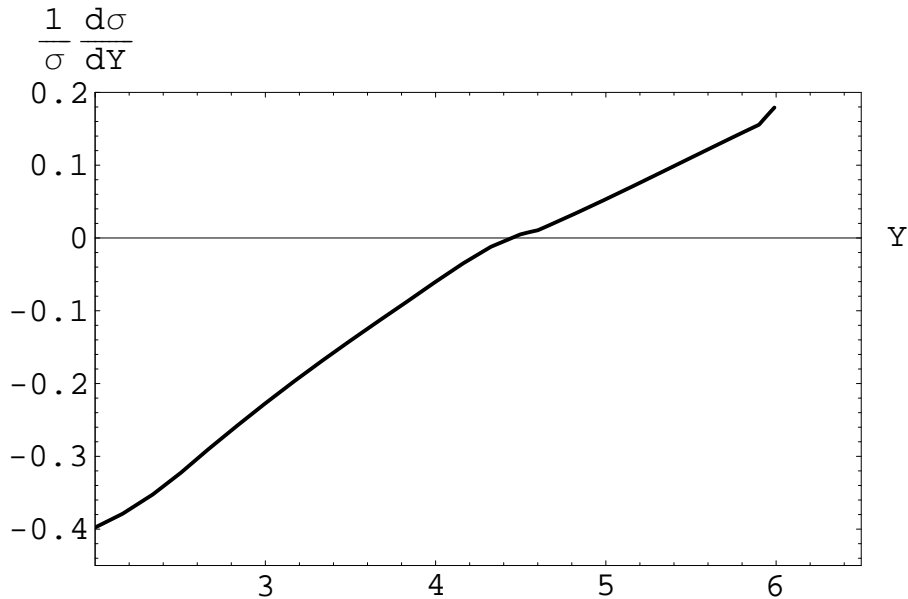}}
\caption[]{Energy dependence of the Pomeron intercept (minus 1) calculated from
the total cross section with the PMS method at $Q^2$=17~GeV$^2$ and $n_f=4$.}
\label{slope}
\end{figure}

From the energy dependence of the NLO BFKL cross section determined through the
PMS method at $Q^2$=17~GeV$^2$ we obtained also the {\em effective} intercept
(minus 1) as a function of the energy. The result is shown in
Fig.~\ref{slope}: it turns out that the intercept grows monotonically in the
energy range considered, in an approximately linear manner. In particular, at
small $Y$ the dynamical intercept is negative since quark box dominates and
its energy behavior mimics a subleading Reggeon; at large $Y$ the perturbative
Pomeron is visible; around $Y\sim 4.5$ the transition between the two regimes
takes place.

\begin{figure}[ht]
\centering
\hspace{-1cm}
{\epsfysize 8cm \epsffile{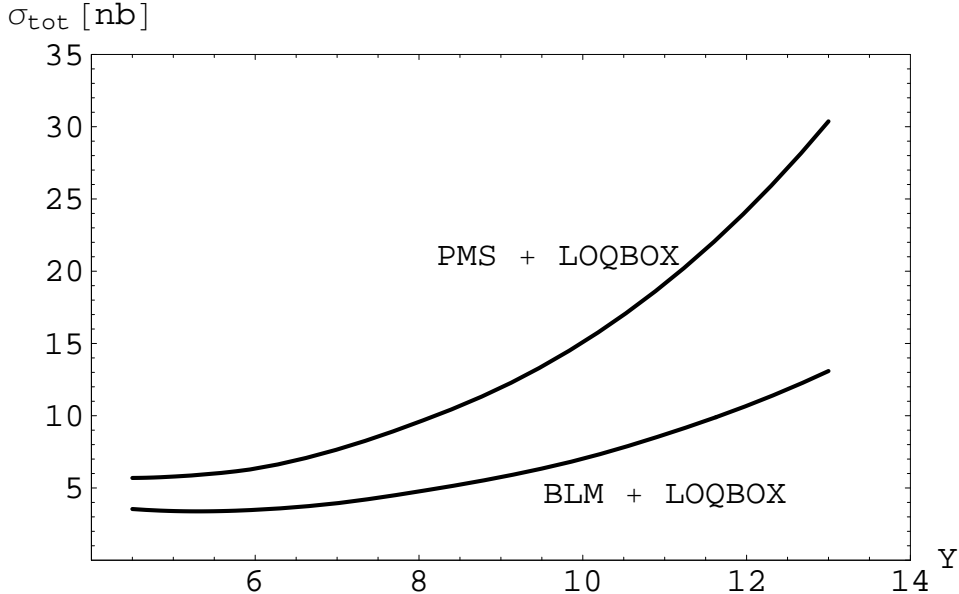}}
\caption[]{Energy dependence of the total cross section for the collision of
two photons with virtualities $Q^2$=20~GeV$^2$ ($n_f$=5) as predicted by the
PMS and BLM methods.}
\label{sigma2}
\end{figure}

In Fig.~\ref{sigma2} we show the $Y$-behavior of the total cross section for
$Q^2$=20~GeV$^2$ ($n_f$=5) in an energy region not explored by past and present
experiments, but relevant for future colliders. We plot here the two curves
obtained in the present work with the PMS and the BLM methods. The
applicability condition for the BFKL resummation, $\bar \alpha_s Y\sim 1$,
corresponds here to $Y\sim 6$; around this energy the deviation between the
PMS and the BLM methods is about $40\%$.
This discrepancy can be taken as an estimate of the systematic uncertainty of
this approach. We observe that our determination from the BLM method is
in quite good agreement with the result of Ref.~\cite{Brodsky:2002ka}
(see Fig.~4 of that paper), obtained for the same kinematics.
The optimal values of the energy scales in the PMS method are similar to those
obtained in the kinematics region studied before ($Q^2$=17~GeV$^2$) for the
lower energies, with a tendency to increase for the higher energies considered.

\section{Conclusions}

In this paper we have presented an estimate of the energy dependence of
the $\gamma^* \gamma^*$ total hadronic cross section in an energy range which
covers LEP2 and future linear colliders. We have used the singlet forward
BFKL Green's function at the next-to-leading order together with forward
$\gamma^* \to \gamma^*$ impact factors {\it at the leading order}. However, we
included in the impact factors the subleading terms required by the
invariance of the full amplitude at the NLA under change of the renormalization
scale $\mu_R$ and of the energy scale $s_0$ entering the BFKL approach.

We have found that,  in spite of the presence of very large NLA corrections,
if suitable methods are used to stabilize the perturbative
series, a smooth curve for the energy behavior of the cross section can be
achieved. We have considered two energy regions: the CERN LEP2 region
and a region possibly reachable by future linear colliders.

Our result in the CERN LEP2 region compares favorably with experimental
data. Systematic effects coming from the optimization procedure are estimated
by the comparison with two different methods. We stress, however, that in
this energy region the role of the neglected NLO corrections to the impact
factors can be relevant; in particular, being these corrections negative, our
prediction certainly overestimates the true  NLO result. Our findings in the
CERN LEP2 region are in agreement with the result of
Ref.~\cite{Brodsky:2002ka}, where for the first
time subleading BFKL effects were considered in the $\gamma^* \gamma^*$ total
hadronic cross section.
Our calculation profits by the experience accumulated in the
last few years with the application of the BFKL approach at subleading level
in the case of the electroproduction of two light vector
mesons~\cite{mesons_1-2,mesons_3}. It can be considered as a refinement of the
analysis of Ref.~\cite{Brodsky:2002ka}, since we apply perturbative series
optimization procedures directly to the process cross section.

The numerical effect of the neglected subleading corrections to the impact
factors cannot be quantified. We expected that it be modest in the second
region of energy considered in this work. Here we believe that our prediction
from the PMS method should be very close to the complete NLA BFKL result.
The final word will be said when the $\gamma^* \gamma^*$ cross section will be
calculated fully in the next-to-leading approximation.

\vspace{1.0cm} \noindent
{\Large \bf Acknowledgment} \vspace{0.5cm}

The work of D.~I. was supported by the grants 08-02-00334-a and
NSh 1027.2008.2.

\end{document}